# Functor is to Lens as Applicative is to Biplate
## Introducing Multiplate

by Russell O'Connor

Department of Computing and Software, McMaster University
Hamilton, Ontario, Canada

*Email:* `roconn@mcmaster.ca`

**Abstract**

This paper gives two new categorical characterisations of lenses: one as a coalgebra of the store comonad, and the other as a monoidal natural transformation on a category of a certain class of coalgebras. The store comonad of the first characterisation can be generalized to a Cartesian store comonad, and the coalgebras of this Cartesian store comonad turn out to be exactly the Biplates of the Uniplate generic programming library. On the other hand, the monoidal natural transformations on functors can be generalized to work on a category of more specific coalgebras. This generalization turns out to be the type of **compos** from the Compos generic programming library. A theorem, originally conjectured by van Laarhoven, proves that these two generalizations are isomorphic, thus the core data types of the Uniplate and Compos libraries supporting generic program on single recursive types are the same. Both the Uniplate and Compos libraries generalize this core functionality to support mutually recursive types in different ways. This paper proposes a third extension to support mutually recursive data types that is as powerful as Compos and as easy to use as Uniplate. This proposal, called Multiplate, only requires rank 3 polymorphism in addition to the normal type class mechanism of Haskell.

## 0 Licence

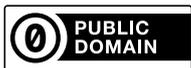

## 1 Introduction

When programming, we often have tree-like data structures that we want to manipulate by pulling out some subtree and replacing it with a new subtree of the same type. This is especially true when manipulating abstract syntax trees where we often want to apply a transformation to all subtrees of a certain type and often these abstract syntax trees are built from mutually recursive data types.

Consider the following small example language [1].

```
data Stm ::= SDecl    Typ Var
          |  SAss     Var Expr
          |  SBlock   [Stm]
          |  SReturn  Expr

data Expr ::= EStm  Stm
           |  EAdd  Expr Expr
           |  EVar  Var
           |  EInt  Int
```

```
data Var := V String
data Typ := TInt
         | TFloat
```

If you want to write a function that given a Stm or an Expr renames all variable names by prepending an underscore then, naively, you would write two mutually recursive functions to traverse the statements and expressions down to the variable names (see Figure 1). These mutually recursive functions must have a case for each constructor, and for every function like this you would have to write another pair of mutually recursive functions.

As the abstract data type grows in complexity, and as the number of such functions grows, the amount of boilerplate code needed also grows; however, all this boilerplate code is essentially the same. Therefore, we would like to find a way to abstract away this common functionality.

```
renameStm  (SDecl t v)    := SDecl t (renameVar v)
renameStm  (SAss v e)     := SAss (renameVar v) (renameExpr e)
renameStm  (SBlock ss)    := SBlock (map renameStm ss)
renameStm  (SReturn e)    := SReturn (renameExpr e)
renameExpr (EStm s)       := EStm (renameStm s)
renameExpr (EAdd e₁ e₂)   := EAdd (renameExpr e₁) (renameExpr e₂)
renameExpr (EVar v)       := EVar (renameVar v)
renameExpr (EInt i)       := EInt i
renameVar  (V s)          := V ('_' : s)
```

**Figure 1.** Naive method of prepending an underscore to all variable names in an abstract syntax tree

Over the last several years many libraries supporting such generic functional programming have appeared. See Rodriguez et al. [9] for a comparison of nine such libraries. Most of these libraries make heavy use of either compile-time or run-time syntax reflection and require compiler support. In this paper, we will only be concerned with the light-weight generic programming libraries such as Uniplate [7] and Compos [1] which are semantics based as opposed to syntax based. With semantics based generic programming one can explicitly define what the substructures of a data type are as opposed to having to deriving this from the syntax of the type declaration. With a semantics based generic programing one can define abstract substructures that are not necessarily syntactical substructures (for example, you can manipulate the coefficients of a sparse matrix as if it were a dense matrix).

In this paper, we will show that two of these competing light-weight libraries, Uniplate and Compos, are using (morally speaking) isomorphic data types for their core functionality on single recursive data types. Each library extends this common core functionality to support mutually recursive data types in different ways. In this paper, we will develop a third library for generic programming on mutually recursive data types called Multiplate that only uses rank 3 polymorphism in addition to the normal type class mechanism of Haskell. With Multiplate, and small amount of initial boilerplate code, we can then write the above rename function as simply:

```
rename := mapFamily (purePlate {var := λ(V s) → pure (V('_' : s))})
renameStm := stm rename¹
renameExpr := expr rename
```

However, before building this library we will first review the most basic structure used for manipulating subexpressions: lenses. The lens structure, also known as a functional reference or an accessor, is defined as a pair of a getter and a setter functions for a structure of type $\alpha$ and a substructure of type $\beta$.

---

1. Actually, in Haskell one would need to write renameStm := runIdentity ∘ stm rename. I am omitting the newtype wrappers and unwrappers in order to clarify the real content of the functions.

$$\textbf{data}\ \mathsf{Lens}\ \alpha\ \beta := \mathsf{Address}\ \left\{\begin{array}{l}\mathsf{get}\ ::\ \alpha \to \beta \\ \mathsf{set}\ ::\ \alpha \to \beta \to \alpha\end{array}\right\}$$

Lenses are particularly nice to work with because they can be composed and because they are first class values, they can also be passed around as parameters. One can use lenses to get, set, and modify a particular substructure of a given structure.

In sections 2.2 and 4.1, we will see two different representations of this lens data structure. This paper gives two novel characterisations of lenses, the first as a coalgebra of the store comonad, and the second as a monoidal natural transformation.

Building upon this foundation, we show in Section 3.2 that the core data type from Uniplate, called Biplate, is morally speaking, a generalization of the first lens representation. We give a novel characterisation of Biplates by showing that they are the coalgebras of the Cartesian store comonad.

On the other hand, we will show in Section 4.2 that the core data type used in Compos is a generalisation of the second representation of a lens. We note that the laws for Compos are also the laws of a monoidal natural transformation. We show that the two core data types of Compos and Uniplate are isomorphic using a theorem originally conjectured by van Laarhoven [13].

Using this new theoretical foundation, we build the Multiplate library. Multiplate uses monoidal natural transformations of a vector of coalgebras as its core data type. By using this new vector approach, Multiplate can support mutually recursive data types as easily as Compos and Uniplate support single recursive data types.

## 1.1 Notation

In this paper I will be using a informal language that is somewhere between Haskell and system $F_\omega$. As in system $F_\omega$, I will explicitly pass type parameters to functions. However, to lower the noise of notation, these parameters will be passed as subscripts and omitted all together when it is clear from context what they should be. In practice, this makes the expressions look like Haskell expressions most of the time.

Another difference between Haskell and my notation is that I will allow class instances to be defined on type synonyms. For example, I define the instances for the identity applicative functor [6] and the composition of applicative functors as follows.[2]

$\textbf{type}\ \mathsf{Id}\ \alpha := \alpha$

$\textbf{instance}\ \mathsf{Functor}\ \mathsf{Id}\ \textbf{where}$
   $\mathsf{fmap}\ f\ x := f\ x$

$\textbf{instance}\ \mathsf{Applicative}\ \mathsf{Id}\ \textbf{where}$
   $\mathsf{pure}\ x := x$
   $f\ \langle * \rangle\ x := f\ x$

$\textbf{type}\ (F \circ G)\ \alpha := F(G\ \alpha)$

$\textbf{instance}\ (\mathsf{Functor}\ F, \mathsf{Functor}\ G) \Rightarrow \mathsf{Functor}\ (F \circ G)\ \textbf{where}$
   $\mathsf{fmap}_{F \circ G}\ f\ x := \mathsf{fmap}_F\ (\mathsf{fmap}_G f)\ x$

$\textbf{instance}\ (\mathsf{Applicative}\ F, \mathsf{Applicative}\ G) \Rightarrow \mathsf{Applicative}\ (F \circ G)\ \textbf{where}$
   $\mathsf{pure}_{F \circ G}\ x := \mathsf{pure}_F\ (\mathsf{pure}_G\ x)$
   $f\ \langle * \rangle_{F \circ G}\ x := (\langle * \rangle_G)\ \langle \$ \rangle_F\ f\ \langle * \rangle_F\ x$

---

2. The $\langle \$ \rangle$ operator is infix notation for the fmap function and the $\langle * \rangle$ operator is infix notation for the ap function. These two operators associate the same way function application does, so $f\langle \$ \rangle x \langle * \rangle y \langle * \rangle z$ stands for $((f\langle \$ \rangle x)\langle * \rangle y)\langle * \rangle z$. This example is a common idiom for lifting a pure 3-ary function $f$ and applying it to three applicative arguments.

This use of type synonyms instances is for presentation purposes only and it is not required. In a Haskell implementation, one would make instances on **newtype** wrappers. Avoiding wrapping and unwrapping **newtype**s makes the presentation here clearer.

## 2   The Store Comonad and Lenses

A comonad is a type constructor class that is dual to the well-known monad class.

```
class Functor w ⇒ Comonad w where
   extract :: w α → α

   duplicate :: w α → w (w α)
   duplicate := extend id

   extend :: (w α → γ) → w α → w γ
   extend f x := fmap f (duplicate x)
```

Here extract is dual to return, duplicate is dual to join, and extend is dual to ($=\!\!\ll$). In analogy with monads, extract and duplicate form a minimal definition of the comonad class, and extract and extend also form a minimal definition of the comonad class. Again, in analogy with monad, if one defines extract and extend first, one can obtain fmap for free by defining it to be liftW.

```
liftW :: Comonad w ⇒ (α → β) → w α → w β
liftW f := extend (f ∘ extract)
```

Comonads are subject to the comonad laws. In addition to the two functor laws, there are five laws that need to be satisfied.

$$
\begin{aligned}
\mathsf{extract} \circ \mathsf{fmap}\ f &= f \circ \mathsf{extract} \\
\mathsf{duplicate} \circ \mathsf{fmap}\ f &= \mathsf{fmap}\ (\mathsf{fmap}\ f) \circ \mathsf{duplicate} \\
\mathsf{extract} \circ \mathsf{duplicate} &= \mathsf{id} \\
\mathsf{fmap}\ \mathsf{extract} \circ \mathsf{duplicate} &= \mathsf{id} \\
\mathsf{fmap}\ \mathsf{duplicate} \circ \mathsf{duplicate} &= \mathsf{duplicate} \circ \mathsf{duplicate}
\end{aligned}
$$

The first two of these laws are the naturality conditions and they come for free [16] (assuming that fmap already satisfies the functor laws). The last three laws of the coherence conditions and they need to be verified for each potential comonad.

### 2.1   The Store Comonad

One of the primary comonads of interest in this paper is the Store comonad, which is dual to the State monad.[3] It is defined as

```
data Store β α := Store { peek :: β → α
                          pos  :: β     }
```

A value of type Store $\beta\alpha$ represents a collection of values of type $\alpha$, where each element of the collection is indexed by a value of $\beta$. There is one element for every $\beta$. The collection is represented by the peek component of type $\beta \to \alpha$. Within the collection there is one special "selected" location. The index for this special location is represented by the pos component of type $\beta$. Figure 2 illustrates what values of this data type look like.

---

3. This dual of the state monad has been given many different names: costate, state-in-context, context, FunArg [10], array. I am proposing yet another name.

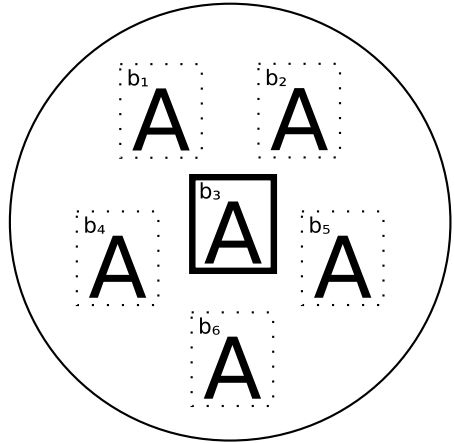

**Figure 2.** Illustration of a value of type Store $\beta\ \alpha$. A collection of various values of type $\alpha$ is denoted by a circle with A's inside. Even though the values can vary, for this illustration I just use A everywhere. Each value A belongs to a location labeled with $b_i$. There is one location for each value of $\beta$ and all locations are occupied. This illustration shows only five locations. Among all the locations, there is one location that is selected, which is indicated by the heavy outlined location. One can think of this location as where a reading head on a disk platter is parked or where a forklift in a warehouse is parked. In this example, the location $b_3$ is selected. The value held in this location is the value returned by extract.

The data type Store $\beta$ forms a comonad for every $\beta$ with the following comonadic operations.

**instance** Functor (Store $\beta$) **where**
  fmap $f$ (Store $v\,b$) := Store $(f \circ v)\,b$

**instance** Comonad (Store $\beta$) **where**
  extract (Store $v\,b$) := $v\,b$
  duplicate (Store $v\,b$) := Store (Store $v$) $b$

The fmap function applies a function to each value at each location. The extract function returns the value held in the selected location. The duplicate function is more interesting. It produces a collection of all the possible selections for the input. This collection is arranged so that a copy of the original collection but with cell $b_i$ selected is put into cell $b_i$. See Figure 3 for an illustration of a result of duplicate. Notice that the original collection ends up placed in the originally selected cell. This property is one of the comonad laws:

$$\text{extract} \circ \text{duplicate} = \text{id}$$

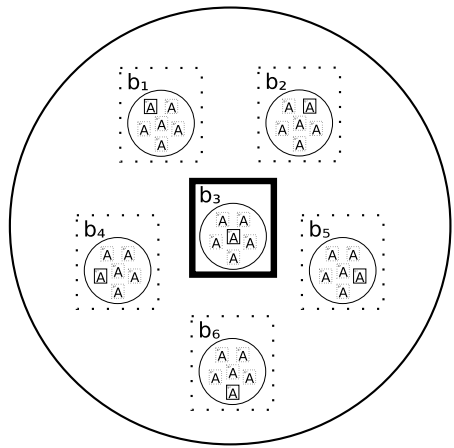

**Figure 3.** An illustration of the result of duplicate applied to the input illustrated in Figure 2.

The function extend takes a given function operating on stores and applies it to all possible selections of the given store.

## 2.2 Lenses

A lens can be represented by the following data type.

**type** Lens $\alpha\,\beta := \alpha \to$ Store $\beta\,\alpha$

The most typical use case for a lens is as a function reference to access a field of a record. For example, suppose we have the following record structure for data for an address book [2]:

**data** Address $:=$ Address $\left\{\begin{array}{ll}\text{phone}\_ & :: \text{PhoneNumber} \\ \text{website}\_ & :: \text{URI}\end{array}\right\}$

We might define address data for Pat as follows:

pat $:=$ Address $\left\{\begin{array}{ll}\text{phone}\_ & := 333\text{-}4444 \\ \text{website}\_ & := \text{http://pat.com/}\end{array}\right\}$

We can make a lens that is a reference to the phone number field of the record.

phone :: Lens Address PhoneNumber
phone $:= \lambda\,address \to$ Store $(\lambda\,newPhone \to address\,\{\text{phone}\_ := newPhone\})\,(\text{phone}\_\ address)$

A similar lens for the website field can also be defined.

The phone lens pairs up the getter and setter functions to access to a single Address into one Store PhoneNumber Address comonadic value. The pos component of the store value is the phone number of the given address. The peek component of the store value is an update function that, given a new phone number, returns a new address with an updated phone number.

A more denotative perspective views this lens as a function that given an address returns a store consisting of all the possible addresses formed by updating the input address with all possible different phone numbers where each possible address is located in the box labeled by the updated phone number. The selected value is the one box labeled with the current phone number and it contains the original address.

A lens is isomorphic to a pair of getter and setter functions

$$\text{Lens}\,\alpha\,\beta \approx (\alpha \to \beta) \times (\alpha \to \beta \to \alpha)$$

The functions that implement this isomorphism are as follows. Given a lens we can retrieve the getter and setter functions.

get :: Lens $\alpha\,\beta \to \alpha \to \beta$
get $l\,a :=$ pos $(l\,a)$

set :: Lens $\alpha\,\beta \to \alpha \to \beta \to \alpha$
set $l\,a :=$ peek $(l\,a)$

Conversely, given getter and setter functions we can build a lens.

lens :: $(\alpha \to \beta) \to (\alpha \to \beta \to \alpha) \to$ Lens $\alpha\,\beta$
lens $gt\,st := \lambda\,a \to$ Store $(st\,a)\,(gt\,a)$

Using the get and set functions we can retrieve and update the fields of our record. The following are examples of accesses the phone field of the pat record:

$$\begin{aligned} \mathsf{get\,phone\,pat} &= 333\text{-}4444 \\ \mathsf{set\,phone\,pat\,}555\text{-}6666 &= \mathsf{Address} \begin{cases} \mathsf{phone\_} &:= 555\text{-}6666 \\ \mathsf{website\_} &:= \text{http://pat.com/} \end{cases} \end{aligned}$$

We expect the get and set functions to satisfy certain laws. Kagawa [4] lists three laws which agree with the laws of a "very well behaved lens" [2].

$$\begin{aligned} \mathsf{get}\,l\,(\mathsf{set}\,l\,s\,b) &= b \\ \mathsf{set}\,l\,s\,(\mathsf{get}\,l\,s) &= s \\ \mathsf{set}\,l\,(\mathsf{set}\,l\,s\,b_1)\,b_2 &= \mathsf{set}\,l\,s\,b_2 \end{aligned} \qquad (1)$$

It is possible to reexpress these laws using the comonadic operations for the store comonad.

$$\begin{aligned} \mathsf{extract} \circ l &= \mathsf{id} \\ \mathsf{fmap}\,l \circ l &= \mathsf{duplicate} \circ l \end{aligned} \qquad (2)$$

A proof that for all $l$ of type $\mathsf{Lens}\,\alpha\,\beta$ (1) holds if and only if (2) holds can be found in the supplementary material [8].

In general, a coalgebra for a functor $F$ is simply a function $f :: A \to FA$ for some type $A$. However, when $W$ is a comonad, we say that $f :: A \to WA$ is a coalgebra for the comonad $W$ when the above two laws are satisfied. This means that lenses are exactly the coalgebras for the store comonad![4]

Lenses can refer to more than just fields of records. They can be used to reference to an element in an array or any substructure of a larger structure, or anything else that satisfies the lens laws above.

Lenses are also composable in the sense that they form a category. For any data structure there is a trivial identity reference from itself to itself. If you have a lens referring to a field of a record and another lens that refers to a field of that field, then those two lenses can be combined into a lens that refers directly to the inner field from the outermost record.

```
idLens :: Lens α α
idLens := Store id

composeLens :: Lens β γ → Lens α β → Lens α γ
l₁ 'composeLens' l₂ := λa → let Store v b := l₂ a in fmap v (l₁ b)
```

In Section 4, we will see a different representation of lenses where lens composition is represented directly by function composition.

### 2.2.1 The Duplicate Lens

For any comonad $W$ and for any type $\alpha$, the duplicate function is a coalgebra for $W$. This means that duplicate is some sort of lens for the store comonad.

```
duplicate :: Lens (Store β α) β
```

---

4. Johnson et. al. shows that lenses are also the algebras of a certain monad over a slice category [3].

We see by this signature that duplicate is a reference to a $\beta$ field of the Store $\beta\alpha$ record. It is not hard to see that this lens is a reference to the pos component of the Store $\beta\alpha$. This lens can be used to get and set the selected location in a store.

## 3　The Cartesian Store Comonad and Biplates

Mitchell and Runciman [7] define their BiplateType, which we will call Biplate, as follows.

**type** Biplate $\alpha\,\beta := \alpha \to ([\beta] \times ([\beta] \to \alpha))$

We can see that this data type as it stands is isomorphic to Lens $\alpha\,[\beta]$; however, Biplates have the constraint that only lists of the same length as the first $[\beta]$ component are accepted by the second $[\beta] \to \alpha$ component. In this sense, Biplates are functional references to multiple substructures at the same time. These substructures can be retrieved and simultaneously updated similar to lenses.

Mitchell and Runciman would really prefer to write the following data type.

**type** Biplate $\alpha\,\beta := \alpha \to \exists n::\mathbb{N}.\,\beta^n \times (\beta^n \to \alpha)$

However, this data type cannot be expressed as such in Haskell, so they are forced to use a data type coarser than they really want. It is too bad that we cannot express such a type in Haskell because $\exists n::\mathbb{N}.\,\beta^n \times (\beta^n \to \alpha)$ forms a comonad similar to the store comonad. Or is there a way to express this type in Haskell?

### 3.1　The Cartesian Store Comonad

One natural way of expressing the type $\exists n::\mathbb{N}.\,\beta^n \times (\beta^n \to \alpha)$ in Haskell is using GADTs with type level natural numbers to create a type of vectors. However, van Laarhoven has defined a type isomorphic to our desired comonad directly using nested data types [13] which works in plain Haskell '98. van Laarhoven called his data type FunList,[5] but I will call it the *Cartesian store comonad*.

**data** CartesianStore $\beta\,\alpha :=$ Unit $\alpha$
　　　　　　　　　　　| Battery (CartesianStore $\beta\,(\beta \to \alpha))\,\beta$

A proof that this CartesianStore $\beta\,\alpha$ data type is isomorphic to $\exists n::\mathbb{N}.\,\beta^n \times (\beta^n \to \alpha)$ can be found in the supplementary material [8]. When thinking about the semantics of the Cartesian store, it is helpful to keep the $\exists n::\mathbb{N}.\,\beta^n \times (\beta^n \to \alpha)$ representation in mind. The dimension of a Cartesian store can be computed by counting the number of Battery constructors.

```
dimension :: CartesianStore β α → ℕ
dimension (Unit _)      := 0
dimension (Battery v _) := succ (dimension v)
```

The Cartesian store data type is similar to the store data type. The difference is that in the Cartesian store data type labels have some extra structure. The items are indexed by a coordinate system of some dimension. Figure 4 illustrates what a value of a Cartesian store of dimension two looks like.

---

5. This is presumably in reference to FunArg which is Uustalu and Vene's [10] name for the store comonad.

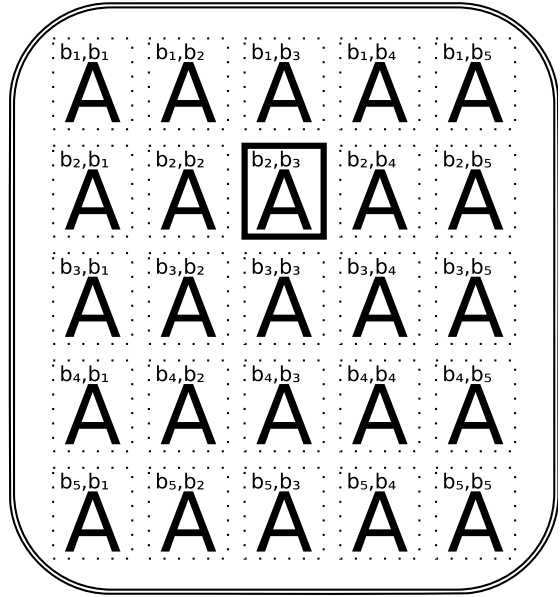

**Figure 4.** An illustration of a value of a Cartesian store comonad. This particular value is two dimensional. The selected location is at $\langle b_2, b_3 \rangle$. Although this figure shows each row and column having a particular order, in a Cartesian store comonad the columns do not necessarily have any particular order within them. On the other hand, the dimensions within the Cartesian store comonad are ordered. The vector used to index has particular first, second, ..., $n^{\text{th}}$ components.

van Laarhoven defines comonadic operations for the Cartesian store as follows:

**instance** Functor (CartesianStore $\beta$) **where**
  fmap $f$ (Unit $a$)      := Unit $(f a)$
  fmap $f$ (Battery $v b$) := Battery (fmap $(f \circ) v) b$

**instance** Comonad (CartesianStore $\beta$) **where**
  extract (Unit $a$)      := $a$
  extract (Battery $v b$) := extract $v b$
  duplicate (Unit $a$)      := Unit (Unit $a$)
  duplicate (Battery $v b$) := Battery (extend Battery $v) b$

The extra structure in the Cartesian store allows us to make an instance of another familiar structure: an applicative functor. Recall that the Applicative class requires two functions, pure and $\langle * \rangle$, in addition to a Functor prerequisite.

**class** Functor $\kappa \Rightarrow$ Applicative $\kappa$ **where**
  pure :: $\alpha \to \kappa \alpha$
  $(\langle * \rangle)$ :: $\kappa (\alpha \to \gamma) \to \kappa \alpha \to \kappa \gamma$

McBride and Paterson give the laws for an applicative functor [6]. van Laarhoven shows that the Cartesian store is an instance of an applicative functor with the following functions [12]:

**instance** Applicative (CartesianStore $\beta$) **where**
  pure := Unit
  $f \langle * \rangle$ (Unit $a$)      := fmap $(\$ a) f$
  $f \langle * \rangle$ (Battery $v b$) := Battery $((\circ) \langle \$ \rangle f \langle * \rangle v) b$

This applicative functor instance will play an important rôle later in Section 4.2. A proof that this is an applicative functor can be found in the supplementary material [8].

A Cartesian store is a generalization of a store. Every store can be transformed into a one dimensional Cartesian store using the singleStore injection below.

```
singleStore :: Store β α → CartesianStore β α
singleStore (Store v b) := Battery (Unit v) b
```

We can also strip off the leading dimension of a Cartesian store (if the dimension is non-zero) as a store, leaving another Cartesian store with dimension one less.

```
stripDimension :: CartesianStore β α → Maybe (Store β α × CartesianStore β α)
stripDimension (Unit a)      := Nothing
stripDimension (Battery v b) := Just (Store (extract v) b, fmap ($ b) v)
```

By iterating stripDimension, we can produce a list of stores that include the results of varying a single dimension of the original Cartesian store while keeping the coordinate of all the other dimensions fixed.

```
stores :: CartesianStore β α → [Store β α]
stores := unfoldr stripDimension
```

Note that only the data along the "axes" around the selected position in the Cartesian store is preserved by stores. The rest of the information in the structure is lost.

## 3.2  Biplates

From the previous section we saw that we can define a comonad in Haskell that captures the invariant that the function component of the result only accepts inputs that are the same length as the data component of the value. A Biplate is thus defined:

```
type Biplate α β := α → CartesianStore β α
```

Once we define Biplates this way, it is natural to require that they be coalgebras of the Cartesian store comonad. As such, they need to satisfy the two laws for coalgebras:

$$\begin{aligned} \text{extract} \circ l &= \text{id} \\ \text{fmap}\, l \circ l &= \text{duplicate} \circ l \end{aligned}$$

This will imply laws about getting and setting values with a Biplate. Neither the above laws nor anything equivalent appear in Mitchell and Runciman's paper [7] but they are implicit in one's understanding of their work.

A Biplate is a generalization of lens because composing with singleStore is an injection from lenses to Biplates. To understand what this injection is doing, recall that a lens denotes a reference to a substructure inside some larger structure while a Biplate allows one to reference an ordered list of zero or more substructures of the same type inside a some larger structure. In this sense we see that a lens is a specialization of Biplates where the number of substructures referenced is exactly one. Thus we can consider Biplates to be a functional multireferences. This generalization of lenses is what van Laarhoven was after in his work.

In Mitchell and Runciman's work [7], canonical Biplates for structures are defined using Haskell's class mechanism. These canonical Biplates reference the maximal subexpressions, or *children*, of the larger structure. With these canonical Biplates, Mitchell and Runciman build a library of fast, lightweight, generic traversal functions for structures.

Biplates also form a category in a similar way to lenses. There is an identity Biplate and a way of composing Biplates.

```
idBiplate :: Biplate α α
idBiplate := Battery (Unit id)
```

```
composeBiplate :: Biplate β γ → Biplate α β → Biplate α γ
composeBiplate o₁ o₂ := f ∘ o₂
  where
    f :: CartesianStore β δ → CartesianStore γ δ
    f (Unit a)     := Unit a
    f (Battery v b) := f v ⟨∗⟩ o₁ b
```

## 4 Polymorphic Representations

Given types $A$ and $B$, another representation of the Store $A\,B$, derived from van Laarhoven's work [11], is

$$\forall \kappa :: \mathsf{Functor}.\, (B \to \kappa\,B) \to \kappa\,A$$

The idea is that the parametricity of values of this type will greatly restrict what values of this data type can be. Since a value of this type has to work for any functor $\kappa$, the only tool available is $\mathsf{fmap}_\kappa$. There is no way for this function to produce a value of $\kappa\,\tau$ for any type $\tau$ except by utilizing its parameter $f :: B \to \kappa\,B$. So a value of this type must internally hold on to a value $b :: B$ and then apply $f$ to it yielding $f\,b :: \kappa\,B$. However, it still needs to produce a value of $\kappa\,A$. If this value also internally holds a value $v :: B \to A$, then it could use $\mathsf{fmap}_\kappa\,v$ to transform $\kappa\,B$ into a $\kappa\,A$. Notice that our value cannot hold internally a value of type $\kappa\,B \to \kappa\,A$, because $\kappa$ is a parameter of the polymorphic function which is not available at definition time.

We see that if we have a $B$ and a $B \to A$, then we can produce such a polymorphic function. These two values are exactly the two components of Store $B\,A$.

```
isoStore₁ :: Store B A → ∀κ :: Functor. (B → κ B) → κ A
isoStore₁ (Store v b) := Λκ → λf → fmap_κ v (f b)
```

It turns out that, using the free theorem [16] for this polymorphic type [15], we can prove that these are effectively the only values that this polymorphic type has. A proof that the types Store $B\,A$ and the polymorphic type $\forall \kappa :: \mathsf{Functor}.\,(B \to \kappa\,B) \to \kappa\,A$ are isomorphic can be found in the supplementary material [8]. The inverse of isoStore₁ is defined below.

```
isoStore₂ :: (∀κ :: Functor. (B → κ B) → κ A) → Store B A
isoStore₂ y := y_(Store B) idLens_B
```

This isomorphism implies following set of isomorphisms.

$$\begin{aligned}
\mathsf{Store}\,B\,A &\approx \forall \kappa :: \mathsf{Functor}.\,(B \to \kappa\,B) \to \kappa\,A \\
\mathsf{Lens}\,A\,B &\approx A \to \forall \kappa :: \mathsf{Functor}.\,(B \to \kappa\,B) \to \kappa\,A \\
\mathsf{Lens}\,A\,B &\approx \forall \kappa :: \mathsf{Functor}.\,(B \to \kappa\,B) \to A \to \kappa\,A
\end{aligned}$$

This last representation of a lens is the van Laarhoven representation of a lens [11], or simply a van Laarhoven lens for short.

The modifier for a van Laarhoven lens is defined by instantiating it at the identity functor, Id, yielding a type $(B \to B) \to (A \to A)$; given a function that modifies $B$, one gets a function that modifies $A$ that works by modifying the particular substructure of type $B$ that is referenced by the lens. The setter of a van Laarhoven lens can be defined by in terms of the modifier.

```
modify :: (∀κ :: Functor. (B → κ B) → A → κ A) → (B → B) → A → A
modify y := y_Id

set :: (∀κ :: Functor. (B → κ B) → A → κ A) → A → B → A
set y a b := modify y (const b) a
```

The getter of a van Laarhoven lens is defined by instantiating it at a constant functor and passing in the identity function (compare this with the setter function which is instantiated at the identity functor and passed a constant function).

**type** Const $\beta\alpha := \beta$

**instance** Functor (Const $\beta$) **where**
  fmap $f\, b := b$

get :: $(\forall \kappa :: \text{Functor}.\, (B \to \kappa B) \to A \to \kappa A) \to A \to B$
get $y := y_{(\text{Const } B)}\, \text{id}_B$

Recall that lenses form a category because there is an identity lens and lenses are composable. van Laarhoven lenses are particularly elegant in this regard, because lens composition is implemented as function composition and the identity lens is implemented as the identity function. A disadvantage of van Laarhoven lenses is it that they require support for rank-2 polymorphism to be used effectively.

## 4.1 Monoidal Natural Transformations

Given this isomorphism between the type of lenses and the polymorphic type of van Laarhoven lenses, one can transport the lens laws through the isomorphism to get the laws for van Laarhoven lenses. However reasoning about van Laarhoven lenses this way is akward. Is there a more natural way of expressing the lens laws for van Laarhoven lenses? The answer is: Yes. However, to see this we have to first build up a few more definitions.

**type** Coalgebra $\alpha\, \kappa := \alpha \to \kappa\, \alpha$

The arguments to Coalgebra are flipped from the usual presentation because in this paper we want to think of Coalgebra $\alpha :: (\star \to \star) \to \star$ as a functor from functors-on-types to types. In particular, if we have a natural transformation between functors on types (I write $\kappa_1 \Rightarrow \kappa_2$ for the type of natural transformations from functor $\kappa_1$ to functor $\kappa_2$), then it can be "mapped" over a coalgebra.

**type** $\kappa_1 \Rightarrow \kappa_2 := \forall \alpha.\, \kappa_1\, \alpha \to \kappa_2\, \alpha$

coalgMap :: $(\kappa_1 \Rightarrow \kappa_2) \to$ Coalgeba $\beta\, \kappa_1 \to$ Coalgebra $\beta\, \kappa_2$
coalgMap $\eta\, c := \eta \circ c$

We can rewrite the type of a van Laarhoven lens as

$l :: \forall \kappa :$ Functor. Coalgebra $B\, \kappa \to$ Coalgebra $A\, \kappa$

and we now see that this is the type of a natural transformation from Coalgebra $B$ to Coalgebra $A$. The parametricity of the type of $l$ gives a free theorem that states that $l$ must be a natural transformation

$$\text{coalgMap } \eta \circ l = l \circ \text{coalgMap } \eta$$

or equivalently

$$\forall c.\, \eta \circ (l\, c) = l\, (\eta \circ c)$$

There is more structure at play here though. Both the category of functors and the category of types are monoidal categories. Types form a monoidal category with $\mathbf{1}$ and $\times$, a.k.a. Cartesian products. On the other hand, functors form a monoidal category with Id and $\circ$, a.k.a. functor composition. Coalgebras preserve this monoidal structure with the following operations.[6]

```
idCoalg :: 1 → Coalgebra α Id
idCoalg () := id

composeCoalg :: (Functor κ₁, Functor κ₂) ⇒ (Coalgebra α κ₁ × Coalgebra α κ₂) →
                                              Coalgebra α (κ₁ ∘ κ₂)

composeCoalg (c₁, c₂) := fmap_{κ₁} c₂ ∘ c₁
```

A proof that these operations satisfy the laws for a monoidal functor can be found in the supplementary material [8]. Given that $\mathsf{Coalgebra}\,A$ and $\mathsf{Coalgebra}\,B$ are both monoidal functors, we can now see that $l$ is actually a monoidal natural transformation.

The laws for a monoidal natural transformation $\mathsf{lens}$ are the following.

$$\begin{aligned} l\,(\mathsf{idCoalg}\,()) &= \mathsf{idCoalg}\,() \\ l\,(\mathsf{composeCoalg}\,(c_1,c_2)) &= \mathsf{composeCoalg}\,(l\,c_1, l\,c_2) \end{aligned} \quad (3)$$

In the supplementary material [8], one can find a proof that these two laws are satisfied exactly when the coalgebra laws for the store comonad are satisfied under the isomorphism. Thus our three sets of laws, (1), (2) and (3) are all equivalent to each other. This means we have three different characterisations of two different representations of a lens structure and its associated laws.

## 4.2 van Laarhoven Biplates

A similar polymorphic representation for $\mathsf{CartesianStore}\,A\,B$ exists:

$$\forall \kappa :: \mathsf{Applicative}.\,(B \to \kappa B) \to \kappa A$$

Just like before, parametricity restricts what these values are able to do. However, since we now know that $\kappa$ is an applicative functor, there are more tools available to us. In particular we can use $\mathsf{pure}$ and $\langle * \rangle$.

For example, if our value internally holds a value $a :: A$, then we can implement this type by ignoring the $f :: B \to \kappa B$ argument and simply return $\mathsf{pure}_\kappa\,a$. Alternatively, if our value is holding a value $b :: B$ and a function $v :: B \to A$, then, we can return $\mathsf{pure}_\kappa\,v\,\langle * \rangle_\kappa\,fb$. Because

$$\mathsf{fmap}_\kappa\,v\,x = \mathsf{pure}_\kappa\,v\,\langle * \rangle_\kappa\,x$$

we see that this case is essentially identical to the lens case. We have further possibilities. Our value could be holding two values $b_1, b_2 :: B$ and a function $v :: B \to B \to A$. In this case we could return $\mathsf{pure}_\kappa\,v\,\langle * \rangle_\kappa\,fb_1\,\langle * \rangle_\kappa\,fb_2$. We could keep adding more and more $B$s and adding more and more parameters to $v$. So essentially, if we have a value of type

$$A + (B \to A) \times B + (B^2 \to A) \times B^2 + \ldots \approx \exists n :: \mathbb{N}.(B^n \to A) \times B^n$$
$$\approx \mathsf{CartesianStore}\,B\,A$$

we can produce a value of the polymorphic type

$$\forall \kappa :: \mathsf{Applicative}.\,(B \to \kappa B) \to \kappa A$$

using the following function.

```
isoCartesianStore₁ :: CartesianStore B A → (∀κ :: Applicative. (B → κB) → κA)
isoCartesianStore₁ (Unit a)       := Λκ → λf → pure_κ a
isoCartesianStore₁ (Battery v b) := Λκ → λf → (isoCartesianStore₁ v)_κ f ⟨*⟩_κ fb
```

---

6. I have written $\mathsf{idCoalg}$ and $\mathsf{composeCoalg}$ this way to better show how the monoidal structure is being preserved. In real code, one would leave out the useless () input to $\mathsf{idCoalg}$ and curry $\mathsf{composeCoalg}$.

These are essentially the only possible values of this type. van Laarhoven conjectured that the type CartesianStore $A\,B$ and the polymorphic type $\forall \kappa :: $ Applicative. $(B \to \kappa B) \to \kappa A$ are isomorphic. I have completed the proof of this isomorphism. The proof can be found in the supplementary material [8]. The inverse to isoCartesianStore$_1$ defined as follows:

isoCartesianStore$_2$ :: $(\forall \kappa :: $ Applicative. $(B \to \kappa B) \to \kappa A) \to$ CartesianStore $B\,A$
isoCartesianStore$_2$ $y := y_{(\text{CartesianStore } B)}$ idBiplate$_B$

This isomorphism implies the following set of isomorphisms.

$$\begin{aligned}
\text{CartesianStore } B\,A &\approx \forall \kappa :: \text{Applicative.} (B \to \kappa B) \to \kappa A \\
\text{Biplate } B\,A &\approx A \to \forall \kappa :: \text{Applicative.} (B \to \kappa B) \to \kappa A \\
\text{Biplate } B\,A &\approx \forall \kappa :: \text{Applicative.} (B \to \kappa B) \to A \to \kappa A
\end{aligned}$$

Just as was the case for lenses, the coalgebra laws for Biplates are equivalent under this isomorphism to the laws for a monoidal natural transformation. The proof of this can be found in the supplementary material [8].

As it turns out, the type $\forall \kappa :: $ Applicative. $(A \to \kappa A) \to A \to \kappa A$ is exactly the type of compos from the Compos library for generic programming [1].

Thus we see that, morally, Uniplate and Compos use isomorphic representations. The only difference is that Compos's type more accurately captures the invariants that Uniplate requires the user to ensure by hand. In particular, the claim from the Uniplate paper [7] that

> [...] the Compos library is unable to replicate either universe or transform from [Uniplate's] library.

is false. Because Uniplate and Compos have isomorphic representations, it must be possible to implement the functionality of Uniplate using Compos. Once one realizes that it is possible, it is not very difficult to implement these functions in Compos. Indeed, in the next section, we will be using a Compos-like representation to implement Multiplate, which includes Uniplate's functionality as a special case.

A van Laarhoven Biplate can directly been seen as a generalization of a van Laarhoven lens. Because every applicative functor is a functor, every van Laarhoven lens is a van Laarhoven Biplate. No conversion function is needed.

## 5 Multiplate

Consider again the van Laarhoven representation of a Biplate:

$$\forall \kappa :: \text{Applicative.} (B \to \kappa B) \to (A \to \kappa A)$$

Here $B$ is the type of substructures of a larger structure of type $A$. Suppose we want to extend the van Laarhoven Biplate type to support references to multiple different types of substructures of $A$. The natural way to do this is to add more parameters to the type. For example, if we want a functional multireference from $A$ to 0 or more substructures of types $B$ and $C$ we would use the type

$$\forall \kappa :: \text{Applicative.} (B \to \kappa B) \to (C \to \kappa C) \to (A \to \kappa A)$$

If $A$, $B$, and $C$ are mutually recursive data types, then we want to consider not only the functional multireference from $A$ to its children of types $A$, $B$, and $C$, but also the functional multireference from $B$ to its children of types $A, B,$ and $C$ and from $C$ to its children of types $A, B,$ and $C$. This means we will want three functional multireferences of types:

$$\begin{aligned}
\forall \kappa :: \text{Applicative.} (A \to \kappa A) \to (B \to \kappa B) \to (C \to \kappa C) \to (A \to \kappa A) \\
\forall \kappa :: \text{Applicative.} (A \to \kappa A) \to (B \to \kappa B) \to (C \to \kappa C) \to (B \to \kappa B) \\
\forall \kappa :: \text{Applicative.} (A \to \kappa A) \to (B \to \kappa B) \to (C \to \kappa C) \to (C \to \kappa C)
\end{aligned}$$

The Compos paper [1] claims that

> Even though these implementations would be identical for all type families, it is difficult to provide generic implementations of them without resorting to multiparameter type classes and functional dependencies since the type of the function tuple will depend on the type family.

Compos supports mutually recursive data types but at the cost of requiring the user to rewrite their data types using GADTs.

Uniplate, with its isomorphic implementation, resorts to multiparameter type classes to support generic programming on mutually recursive data types without the need for functional dependencies by introducing Biplates. However, Uniplate's support for mutually recursive data types is limited to dealing with one pair of parent-child types at a time. For instance, it is not possible to update two types of children in one traversal of a parent type.

In this section, we propose an alternative method of supporting mutually recursive data types that does not require GADTs, nor rewriting one's data types, nor does it require multiparameter type classes. However, we will make use of rank 3 polymorphism.

The key observation is that, instead of creating three nearly identical functional multireference types, we can combine them into one "matrix transformation" that operates on a "vector" of coalgebras. To begin, we define a record type parametrized by applicative functions as follows.

$$\mathbf{data}\ \mathsf{P}\kappa := \left\{ \begin{array}{ll} \mathsf{coalgA} :: & A \to \kappa A \\ \mathsf{coalgB} :: & B \to \kappa B \\ \mathsf{coalgC} :: & C \to \kappa C \end{array} \right\}$$

Then we can write one type that incorporates all three of our previous types.

$$\forall \kappa :: \mathsf{Applicative}.\ \mathsf{P}\ \kappa \to \mathsf{P}\ \kappa.$$

Given the record type P, we can provide generic implementations of traversal operations. To see how to accomplish this, we will first write non-generic implementation of this for an example data type, then we will see how to abstract out the generic components.

## 5.1 Mutually Recursive Data Types

Recall the small language from the Introduction.

```
data Stm := SDecl    Typ Var
         |  SAss     Var Expr
         |  SBlock   [Stm]
         |  SReturn  Expr
data Expr := EStm Stm
          |  EAdd  Expr Expr
          |  EVar  Var
          |  EInt  Int
data Var := V String
data Typ := TInt
        |  TFloat
```

For our implementation, we will need a record type with a field for a coalgebra for each of these four types from that small language. This record is parametrized by an applicative functor. We will call such a record a *plate*.

$$\mathbf{data}\ \mathsf{Plate}\ \kappa := \mathsf{Plate} \left\{ \begin{array}{lll} \mathsf{stm} & :: \mathsf{Stm} & \to \kappa\ \mathsf{Stm} \\ \mathsf{expr} & :: \mathsf{Expr} & \to \kappa\ \mathsf{Expr} \\ \mathsf{var} & :: \mathsf{Var} & \to \kappa\ \mathsf{Var} \\ \mathsf{typ} & :: \mathsf{Typ} & \to \kappa\ \mathsf{Typ} \end{array} \right\}$$

We first provide the functional multireference that defines the reference to the children of each of these data types. We call this functional multireference multiplate and it is defined in Figure 5.

$$
\begin{aligned}
&\mathsf{multiplate} :: \mathsf{Applicative}\,\kappa \Rightarrow \mathsf{Plate}\,\kappa \to \mathsf{Plate}\,\kappa \\
&\mathsf{multiplate}\,p := \mathsf{Plate} \left\{ \begin{array}{ll} \mathsf{stm} & := \mathsf{buildStm} \\ \mathsf{expr} & := \mathsf{buildExpr} \\ \mathsf{var} & := \mathsf{buildVar} \\ \mathsf{typ} & := \mathsf{buildTyp} \end{array} \right\}
\end{aligned}
$$

where
 buildStm (SDecl $t\,v$)  := SDecl  $\langle \$ \rangle$ typ $p\,t$    $\langle * \rangle$ var $p\,v$
 buildStm (SAss $v\,e$)  := SAss  $\langle \$ \rangle$ var $p\,v$    $\langle * \rangle$ expr $p\,e$
 buildStm (SBlock $ss$)  := SBlock  $\langle \$ \rangle$ traverse (stm $p$) $ss$
 buildStm (SReturn $e$)  := SReturn  $\langle \$ \rangle$ expr $p\,e$
 buildExpr (EStm $s$)  := EStm  $\langle \$ \rangle$ stm $p\,s$
 buildExpr (EAdd $e_1\,e_2$) := EAdd  $\langle \$ \rangle$ expr $p\,e_1$    $\langle * \rangle$ expr $p\,e_2$
 buildExpr (EVar $v$)  := EVar  $\langle \$ \rangle$ var $p\,v$
 buildExpr $x$  := pure $x$
 buildVar $x$  := pure $x$
 buildTyp $x$  := pure $x$

**Figure 5.** The multiplate function for a little language taken from the Compos paper [1].

Notice that EInt $i$ has no children (at least no children of type Stm, Expr, Var, or Typ) and in this case we simply return the input wrapped in pure and similarly for values of type Var and Typ. Also notice that SBlock's children are held in a list. We use the fact that the list container is Traversable in order to collect them.

Using multiplate, we can recursively define a collection of rename functions, one for each type in our mutually recursive collection of types, that prefixes each variable with an underscore.

$$
\begin{aligned}
&\mathsf{rename} :: \mathsf{Plate}\,\mathsf{Id} \\
&\mathsf{rename} := \mathsf{Plate} \left\{ \begin{array}{ll} \mathsf{stm} & := \mathsf{stm}\,(\mathsf{multiplate}\,\mathsf{rename}) \\ \mathsf{expr} & := \mathsf{expr}\,(\mathsf{multiplate}\,\mathsf{rename}) \\ \mathsf{var} & := \lambda(\mathsf{V}\,s) \to \mathsf{pure}_{\mathsf{Id}}\,(\mathsf{V}\,(\text{'\_':}s)) \\ \mathsf{typ} & := \mathsf{typ}\,(\mathsf{multiplate}\,\mathsf{rename}) \end{array} \right\}
\end{aligned}
$$

The recursive calls to multiplate rename will cause all of the descendants of statements and expressions to be renamed. Each field of this plate will rename variables for a different type. For example, stm rename :: Stm $\to$ Stm, can be used to rename variables in statements, and expr rename :: Expr $\to$ Expr, can be used to rename variables in expressions.

Above, we used the method from Compos to create the rename function. Uniplate improves upon this by defining generic traversal function that will recursively apply a given transformation function bottom-up. Since Uniplate is isomorphic to Compos, it must be possible to write this generic traversal function in the Compos representation.

mapFamily[7] :: Plate Id $\to$ Plate Id
mapFamily $p$ := $p$ `composePlateId` multiplate (mapFamily $p$)
 **where**
  composePlateId :: Plate Id $\to$ Plate Id $\to$ Plate Id

---

 7. These function names are taken from `http://www-ps.informatik.uni-kiel.de/~sebf/projects/traversal.html`.

$$p_1 \text{ `composePlateId` } p_2 := \text{Plate} \begin{Bmatrix} \text{stm} & := & \text{stm} & p_1 \circ \text{stm} & p_2 \\ \text{expr} & := & \text{expr} & p_1 \circ \text{expr} & p_2 \\ \text{var} & := & \text{var} & p_1 \circ \text{var} & p_2 \\ \text{typ} & := & \text{typ} & p_1 \circ \text{typ} & p_2 \end{Bmatrix}$$

We can use mapFamily to define rename by applying it to a plate that only renames top-level variables. To make such a plate it is easiest to start with a generic pure plate that does nothing.

purePlate :: Applicative $\kappa \Rightarrow$ Plate $\kappa$

$$\text{purePlate} := \text{Plate} \begin{Bmatrix} \text{stm} & := & \text{pure} \\ \text{expr} & := & \text{pure} \\ \text{var} & := & \text{pure} \\ \text{typ} & := & \text{pure} \end{Bmatrix}$$

Now we can override this pure plate and easily obtain a plate that only renames top-level variables which we pass to mapFamily to rename all variables.

rename := mapFamily (purePlate {var := renameVar})
  **where**
    renameVar (V $s$) := pure (V('_' : $s$))

Notice that now our definition only mentions the types and the cases that we are interested in. The rest of the types and cases are handled generically.

Reviewing our definition of mapFamily, we see that we can generalize composePlateId to work over any monad by using the Kleisli composition operator.

kleisliComposePlate :: Monad $m \Rightarrow$ Plate $m \to$ Plate $m \to$ Plate $m$

$$p_1 \text{ `kleisliComposePlate` } p_2 := \text{Plate} \begin{Bmatrix} \text{stm} & := & \text{stm} & p_1 \lll \text{stm} & p_2 \\ \text{expr} & := & \text{expr} & p_1 \lll \text{expr} & p_2 \\ \text{var} & := & \text{var} & p_1 \lll \text{var} & p_2 \\ \text{typ} & := & \text{typ} & p_1 \lll \text{typ} & p_2 \end{Bmatrix}$$

Thus our definition of mapFamily generalizes to arbitrary monads. This yields mapFamilyM.

mapFamilyM :: Monad $m \Rightarrow$ Plate $m \to$ Plate $m$
mapFamilyM $p := p$ `kleisliComposePlate` multiplate (mapFamilyM $p$)

## 5.2 Type-Generic Generic Functions for Mutually Recursive Types

The functions in the previous section can be used to build generic traversal functions for the specific mutually recursive data type we defined in that section. However, we would like to define these generic functions generically for all mutually recursive data types. To this end we will create a type class for plates.

For a prospective plate P we will require an instance of the multiplate function having type

multiplate :: Applicative $\kappa \Rightarrow$ P $\kappa \to$ P $\kappa$

We also need a generic way of building plates. Notice that for each of our generic functions purePlate, kleisliComposePlate and mapFamilyM, each field of the record is built in a uniform way. Therefore, if we are given a polymorphic function of type $\forall \alpha. \alpha \to \kappa \alpha$ we can build an arbitrary plate P $\kappa$ for an arbitrary applicative functor $\kappa$. However, our generic functions also need projection function of the field being build. A projection function for a plate P has type Projector P $\alpha$ defined below.

**type** Projector $\rho \alpha := \forall \kappa. \rho \kappa \to \alpha \to \kappa \alpha$

We can pass the projector for the field we are building to a generic builder and have the generic builder create each field of our record. To this end we require the user to provide a second function for each prospective plate.

mkPlate :: $(\forall \alpha. \text{Projector } p\,\alpha \to \alpha \to \kappa\,\alpha) \to p\,\kappa$

Putting these two functions together we form the Multiplate class.

**class** Multiplate $\rho$ **where**
   multiplate :: Applicative $\kappa \Rightarrow \rho\,\kappa \to \rho\,\kappa$
   mkPlate   :: $(\forall \alpha. \text{Projector } \rho\,\alpha \to \alpha \to \kappa\,\alpha) \to \rho\,\kappa$

For our example Plate, the multiplate function is the function defined in the previous section. The mkPlate function for Plate is simply

$$\text{mkPlate } build := \text{Plate} \begin{Bmatrix} \text{stm} & := & build\,\text{stm} \\ \text{expr} & := & build\,\text{expr} \\ \text{var} & := & build\,\text{var} \\ \text{typ} & := & build\,\text{typ} \end{Bmatrix}$$

Now we can define purePlate, kleisliComposePlate, and mapFamilyM and more, generically for all instances of Multiplate.

purePlate :: (Multiplate $\rho$, Applicative $\kappa$) $\Rightarrow \rho\,\kappa$
purePlate := mkPlate (const pure)

idPlate :: Multiplate $\rho \Rightarrow \rho\,\text{Id}$
idPlate := purePlate

mapPlate :: $\forall \rho\,\kappa_1\,\kappa_2.$ Multiplate $\rho \Rightarrow (\forall \gamma. \kappa_1\,\gamma \to \kappa_2\,\gamma) \to \rho\,\kappa_1 \to \rho\,\kappa_2$
mapPlate $\eta\,p$ := mkPlate build
  **where**
    build :: Projector $\rho\,\alpha \to \alpha \to \kappa_2\,\alpha$
    build $\pi := \eta \circ \pi\,p$

composePlate :: $\forall \rho\,\kappa_1\,\kappa_2.$ (Multiplate $\rho$, Applicative $\kappa_1$, Applicative $\kappa_2$) $\Rightarrow \rho\,\kappa_1 \to \rho\,\kappa_2 \to \rho\,(\kappa_2 \circ \kappa_1)$
$p_1$ 'composePlate' $p_2$ := mkPlate build
  **where**
    build :: Projector $\rho\,\alpha \to \alpha \to \kappa_2\,(\kappa_1\,\alpha)$
    build $\pi := \text{fmap}_{\kappa_2}\,(\pi\,p_1) \circ \pi\,p_2$

kleisliComposePlate :: (Multiplate $\rho$, Monad $m$) $\Rightarrow \rho\,m \to \rho\,m \to \rho\,m$
$p_1$ 'kleisliComposePlate' $p_2$ := mapPlate join ($p_1$ 'composePlate' $p_2$)

mapFamilyM :: (Multiplate $\rho$, Monad $m$) $\Rightarrow \rho\,m \to \rho\,m$
mapFamilyM $p := p$ 'kleisliComposePlate' multiplate (mapFamilyM $p$)

Here we define kleisliComposePlate in terms of the more general function composePlate.

We can also write code for generic folding over structures in the same way the Compos does. We define appendPlate which combines plates over the applicative constant functor on a monoid.

**instance** (Monoid $o$) $\Rightarrow$ Applicative (Const $o$) **where**
   pure $x := 1_o$
   $f \langle * \rangle\, x := f *_o x$

appendPlate :: $\forall \rho\,o.$ (Multiplate $\rho$, Monoid $o$) $\Rightarrow \rho\,(\text{Const } o) \to \rho\,(\text{Const } o) \to \rho\,(\text{Const } o)$
$p_1$ 'appendPlate' $p_2$ := mkPlate build

    **where**
        build :: Projector $\rho\,\alpha \rightarrow \alpha \rightarrow m\,\alpha$
        build $\pi\,a := \pi\,p_1\,a \langle * \pi\,p_2\,a$

preorderFold :: (Multiplate $\rho$, Monoid $o$) $\Rightarrow \rho\,(\text{Const}\,o) \rightarrow \rho\,(\text{Const}\,o)$
preorderFold $p := p$ 'appendPlate' multiplate (preorderFold $p$)

postorderFold :: (Multiplate $\rho$, Monoid $o$) $\Rightarrow \rho\,(\text{Const}\,o) \rightarrow \rho\,(\text{Const}\,o)$
postorderFold $p :=$ multiplate (postrderFold $p$) 'appendPlate' $p$

Given a plate $p$ that describes how to convert each type of data into a monoid type $o$, the preorderFold and postorderFold functions create a plate of functions that traverses all the descendants of the inputs and combine the results using the monoid operation.

The function mkPlate is a rank 3 polymorphic function. This is because Projector $\rho\,\alpha$ is a type polymorphic over all type constructors $\kappa$, not just the $\kappa$ bound in the type of mkPlate. It is important that build's projection function parameter is polymorphic because it gets instantiated to different type constructors for some generic functions.

### 5.2.1 Multiplate Laws

Although this library was built to support plates that define one field for each type in a mutually recursive set of types, the resulting library is general enough to support any plate structure that defines multiplate and mkPlate, so long as it satisfies the multiplate laws:

$$\begin{aligned}\text{multiplate idPlate} &= \text{idPlate} \\ \text{multiplate (composePlate } p_1\,p_2) &= \text{composePlate (multiplate } p_1)\,(\text{multiplate } p_2)\end{aligned}$$

The multiplate laws simply state that multiplate is a monoidal natural transformation and are analogous to the laws for van Laarhoven lens that we found in Section 4.1. This means that users are free to define multiplate instances however they choose, so long as these two laws are satisfied. For example, Visscher shows how to define a plate with one field per constructor of the mutually recursive data types [14]. Visscher's definition of plate turns out to be exactly the type MPreserve from "Dealing with Large Bananas" [5]. The types Preserve and Unify can are instances of MPreserve at the identity and constant functors respectively. Thus Multiplate captures much of the functionality of Large Bananas as well as Compos and Uniplate.

# 6 Conclusions and Future Work

In this paper, we defined the store comonad as the dual of the state monad and saw that the coalgebras of this comonad exactly characterize lenses. We define the Cartesian store comonad as a generalization of the store comonad and noted that it was also an applicative functor. We saw that the coalgebras of this comonad exactly characterizes the Biplates from the Uniplate library [7]. After this we defined van Laarhoven representations of both lenses and Biplates. van Laarhoven lenses are monoidal natural transformations of coalgebras that are polymorphic over functors, while van Laarhoven Biplates are monoidal natural transformations of coalgebras that are polymorphic over applicative functors. This completes the analogy given in the title of this paper: Functors are to Lenses as Applicative is to Biplate.

Using this theory we derived a new method of doing generic traversals over mutually recursive data types. This requires the user to define a type, called a plate, for a vector of coalgebras with one field for each type in the mutually recursive set of types. Then the user creates an instance of the Multiplate class for this plate. Once this is defined, all the generic traversal functions of the Multiplate library are available. This paper describes only the most important functions from the Multiplate library.[8]

---

8. `http://hackage.haskell.org/package/multiplate/`

The Multiplate library has a number of advantages over other light-weight generic programming libraries. As noted before, Compos requires the user rewrite his data type as a GADT, effectively reducing a mutually recursive data type into a single recursive data type. Multiplate works with the user's existing data type. The Biplate mechanism of Uniplate only allows you to work with pairs of data types of one's mutually recursive set at a time. Thus with the Uniplate library you cannot easily do traversals of a structure that simultaneously modifies two different data types at the same time. With Multiplate, there is no problem creating a plate that modifies two different data types at the same time. The Multirec library [17] requires the user to build an isomorphism between one's mutually recursive data type and another data type built out of higher order fixpoint combinators and makes extensive use of GADTs and type families. I believe the definitions for Plates and the Multiplate instances required to use Multiplate are much simpler, and the use of rank 3 polymorphism is a milder requirement for the compiler.

The Multiplate class is currently a little unsatisfying. It would be more natural to define the primitive functions as multiplate, idPlate, composePlate, and mapPlate, which is all that is required to state the Multiplate laws. Although purePlate and kleisliComposePlate can be defined in terms of those primitive functions, I do not know how to define appendPlate. Also, although mkPlate is sufficient to define all the functions needed, it seems unnecessarily strong. Further research is needed to refine the Multiplate class. I suspect that the mkPlate function can be eliminated thus dropping the requirement for rank 3 polymorphism down to rank 2 polymorphism.

It seems natural to extend the title's analogy from Applicative to Monad; however the class of functors that are monads is not closed under composition. Thus monads do not have the necessary monoidal structure. On the other hand, Alternative functors are closed under composition. It may be fruitful to study how this analogy might extend to Alternative.

## Acknowledgements

I would like to thank Jacques Carettes, who I discussed this work with on many occasions, for his helpful input. This work grew out of our work with Bill Farmer on the Mathscheme project[9] where we wanted to implement generic traversals for our data structures. I would also like to thank Edward Kmett for his discussions with me. Finally, I would like to thank all the reviewers from ICFP and WGP 2011 who gave me many helpful suggestions. This document has been produced using GNU T<sub>E</sub>X<sub>MACS</sub>.[10]

---

9. `http://www.cas.mcmaster.ca/research/mathscheme/`

10. `http://www.texmacs.org/`